%% file: ifacconf.tex
\documentclass{ifacconf}

\counterwithin*{section}{part}

\usepackage{graphicx}      %
\usepackage{natbib}        %
\usepackage{amsmath,amssymb}
\usepackage{pgfplots}
\pgfplotsset{compat=newest}
\usepackage{bbm}
\usepackage{mathtools}
\usepackage{dsfont}
\usepackage{aligned-overset} %
\usepackage{xcolor}

\usepgfplotslibrary{external}
\usepackage{cleveref}	%
\crefname{equation}{}{}
\crefname{thm}{Theorem}{Theorems}

\definecolor{ceruleanblue}{rgb}{0.16, 0.32, 0.75}
\definecolor{tangelo}{rgb}{0.98, 0.3, 0.0}

\newcommand{\TTT}{\ensuremath{T_{\mathrm{TT}}}}
\newcommand{\TET}{\ensuremath{T_{\mathrm{ET}}}}
\newcommand{\E}[1]{\ensuremath{\mathbb{E}\!\left\lbrack#1\right\rbrack}}
\newcommand{\Prob}{\ensuremath{\mathbb{P}}}
\newcommand{\dbar}[1]{\ensuremath{\bar{\bar{#1}}}}
\newcommand*\diff{\mathop{}\!\mathrm{d}}

\begin{document}
\begin{frontmatter}

	\title{Shared Network Effects in Time- versus Event-Triggered Consensus of a Single-Integrator Multi-Agent System \thanksref{footnoteinfo}}

	\thanks[footnoteinfo]{F. Allgöwer thanks the German Research Foundation (DFG) for support of this work within grant AL 316/13-2 and the German Excellence Strategy under grant EXC-2075 - 285825138; 390740016.}

	\author[First]{David Meister}
	\author[Second]{Frank Dürr}
	\author[First]{Frank Allgöwer}

	\address[First]{University of Stuttgart, Institute for Systems Theory and Automatic Control, Germany
		(e-mail: \{meister,allgower\}@ist.uni-stuttgart.de)}
	\address[Second]{University of Stuttgart, Institute for Parallel and Distributed Systems, Germany
		(e-mail: frank.duerr@ipvs.uni-stuttgart.de)}

	\begin{abstract}	%
		Event-triggered control has the potential to provide a similar performance level as time-triggered (periodic) control while triggering events less frequently.
		It therefore appears intuitive that it is also a viable approach for distributed systems to save scarce shared network resources used for inter-agent communication.
		While this motivation is commonly used also for multi-agent systems, a theoretical analysis of the impact of network effects on the performance of event- and time-triggered control for such distributed systems is currently missing.
		With this paper, we contrast event- and time-triggered control performance for a single-integrator consensus problem under consideration of a shared communication medium.
		We therefore incorporate transmission delays and packet loss in our analysis and compare the triggering scheme performance under two simple medium access control protocols.
		We find that network effects can degrade the performance of event-triggered control beyond the performance level of time-triggered control for the same average triggering rate if the network is used intensively.
		Moreover, the performance advantage of event-triggered control shrinks with an increasing number of agents and is even lost for sufficiently large networks in the considered setup.
	\end{abstract}

	\begin{keyword}
		Event-triggered control, Multi-agent systems, Control over networks, Consensus, Control with data loss.
	\end{keyword}

\end{frontmatter}

\section{Introduction}

\cite{Astrom2002} have demonstrated that event-triggered control (ETC) can outperform time-triggered control (TTC) for a single-loop single-integrator system given equal average triggering rates.
While TTC samples periodically, ETC initiates an event when a designed triggering condition is fulfilled.
\citeauthor{Astrom2002}'s finding gives rise to the conjecture that ETC has the potential to save scarce shared network resources while achieving the same performance level as TTC.
This motivation is not only popular in the area of networked control systems (NCS), e.g., \cite{Henningsson2008} and \cite{Heemels2008}, but also for multi-agent systems (MAS), e.g., \cite{Seyboth2013} and \cite{Nowzari2019}.
To differentiate clearly between NCS that are only coupled through their use of a shared communication medium and MAS in which the agents additionally cooperate on a common goal, we refer to the former as non-cooperative NCS throughout this paper.

While intuitively plausible at first sight, the conjecture above can be falsified in some relevant settings.
As a starting point to examine the conjecture more closely, \cite{Rabi2009} extend the analysis of \cite{Astrom2002} incorporating packet loss effects on a shared network of multiple non-cooperative single-integrator systems.
\cite{Blind2011} and \cite{Blind2011a} additionally consider transmission delay effects and derive packet loss and delay quantities based on two medium access control (MAC) protocols.
These results are complemented by \cite{Blind2013} with the analysis of further MAC protocols in the same setting and their comparison for varying network loads.
The authors thereby demonstrate that network effects can degrade the performance of ETC compared to TTC due to higher packet loss probabilities or transmission delays, depending on the MAC protocol.
Considering the interplay between triggering scheme and network therefore appears crucial for the design of triggering schemes in the non-cooperative NCS case.

One approach to extend the analysis from \cite{Astrom2002} to more general system classes has been the introduction of the notion of consistency for ETC by \cite{Antunes2016}.
An ETC scheme is called consistent if it provides a better performance level than the TTC scheme with the same average triggering rate.
\cite{Balaghiinaloo2022} leverage the consistency framework to evaluate the performance implications of network effects for stochastic ETC and TTC schemes in the case of non-cooperative linear discrete-time systems.
They propose an ETC scheme that is provably consistent for the considered system class and network setting even when some network effects are incorporated.

While performance comparisons of ETC and TTC schemes including network considerations have thus been studied in various works, a respective analysis is still missing for MAS settings.
Therefore, our contribution is to extend our previous theoretical evaluation in \cite{Meister2022,Meister2023}, which does not take network effects into account, by the consideration of shared network effects, namely packet loss and transmission delays.
Similarly to \cite{Blind2013}, we derive packet loss probabilities and expected transmission delays based on two MAC protocols in order to evaluate performance levels of ETC and TTC for various network loads in our single-integrator MAS setup.
We show that ETC performance degradation due to network effects can outweigh its general performance advantage compared to TTC in the considered setup, depending on the network load.
In addition, we demonstrate that for sufficiently large numbers of agents, TTC provides lower network loads than ETC for all attainable performance levels.
Consequently, we exemplify that network effects can have similar performance implications in a MAS setting as in the respective non-cooperative NCS setting while their impact might be even more severe and additional impact factors such as the number of agents might need to be taken into account.

Our paper has the following structure:
In Section~\ref{sec:setup}, we introduce the considered problem setup.
Subsequently, we present our theoretical results in Sections~\ref{sec:ttc_etc} and \ref{sec:mac} where we first derive cost expressions for TTC and ETC and then, quantify packet loss probabilities and expected transmission delays for the considered MAC protocols in order to contrast the derived cost expressions.
We complement our findings with a numerical evaluation of the cost in Section~\ref{sec:sim} and conclude this work in Section~\ref{sec:conclusion}.

\section{Setup and Problem Formulation}\label{sec:setup}

We consider a very similar problem setup as in \cite{Meister2022} but take a network-oriented perspective and include network effects, namely transmission delays and packet loss, in our analysis.
We assume that all agents are connected to a shared communication network and, thus, can communicate with any other agent in the network.
This can be abstracted as an all-to-all communication topology with $N$ nodes representing the agents.

All agents adhere to perturbed single-integrator dynamics
\begin{equation}\label{eq:agent}
	\mathrm{d}x_i = u_i \mathrm{d}t + \mathrm{d}v_i.
\end{equation}
Let the agents start in consensus, i.e., the initial states are $x_i(0)=0$ for all $i\in \lbrace 1,\dots,N\rbrace$.
In addition, let $v_i(t)$ and $u_i(t)$ refer to a standard Brownian motion and the control input, respectively.

Furthermore, we presume that the agents trigger discrete transmission events while being able to monitor their own state continuously.
As explained in the introduction, we aim at comparing TTC and ETC schemes for triggering transmissions.
For this comparison, we consider the cost
\begin{equation}\label{eq:cost}
	J \coloneqq \limsup_{M\rightarrow\infty} \frac{1}{M} \int_{0}^{M} \E{x^\top Lx} \diff t,
\end{equation}
as a performance measure where $x=\lbrack x_1(t), \dots, x_N(t)\rbrack^\top$ and $L$ is the Laplacian of the all-to-all communication graph.
The cost functional quantifies the long-term average of the quadratic deviation from consensus.

An impulsive control input
\begin{equation}\label{eq:input}
	u_i(t) = \sum_{j\in\mathcal{N}_i} \sum_{k\in\mathcal{A}^j_i} \delta(t-t^j_{k}-d^j_{k})(x_j(t^j_{k})-x_i(t^j_{k})),
\end{equation}
is utilized to control the agents.
Let $\mathcal{N}_i = \lbrace1,\dots,N\rbrace\backslash\lbrace i\rbrace$ be the set of neighbors of agent $i$ and $\mathcal{A}^j_i$ be the set of indices $k$ for arrived packets at agent $i$ sent by agent $j$.
In addition, $\delta(\cdot)$ refers to the Dirac impulse.
Furthermore, $t^j_{k}$ and $d^j_k$ denote the transmission time instant and transmission delay of packet $k$ sent by agent $j$, respectively.
Note that we can utilize the state $x_i(t^j_{k})$ in $u_i(t)$ since every agent can continuously monitor its own state.
Thus, computation of the input $u_i(t)$ only requires local and transmitted state information.
Consequently, transmitting an agent's state to all other agents allows for a reset of the MAS to consensus.
Such a reset is only performed if the broadcasted packet arrives and the input is applied after a transmission delay $d^j_{k}$.
Note that the latter point implies that the system does not achieve exact consensus due to the outdated state information used for the input computation.
Between such consensus resets, the agents behave according to standard Brownian motions.

Let us introduce three core assumptions to determine the mechanisms behind packet loss and transmission delays within this work:

\begin{assum}\label{assum:transmission_time}
	Each packet requires a constant transmission time $\tau>0$ in order to travel from the transmitter to the receivers.
	The transmission time is the same for all packets regardless of the transmitter or receiver.
\end{assum}

\begin{assum}\label{assum:collision}
	The shared network behaves like a single communication channel:
	If two or more packets are transmitted via the shared medium at the same time, they collide and are lost.
\end{assum}

\begin{assum}\label{assum:all_or_nothing}
	A packet is either transmitted successfully to all neighboring agents or lost.
\end{assum}

Note that while the transmission time $\tau$ of a packet is constant according to Assumption~\ref{assum:transmission_time}, the transmission delay $d^j_k$ also depends on the deployed MAC protocol. %
Due to Assumption~\ref{assum:all_or_nothing}, we can neglect the subscript for the receiving agent $i$ in $\mathcal{A}^j_i$ and simply write $\mathcal{A}^j$ for the set of indices of arrived packets sent by agent $j$.

Lastly, we utilize the following notation regarding the series of transmission events in this paper:
We have already introduced the series of triggering time instants $(t^j_k)_{k\in\mathbb{N}}$ for agent $j\in\lbrace1,\dots,N\rbrace$.
Moreover, we will also refer to the event series of the complete MAS with the notation $(t_k)_{k\in\mathbb{N}}$.
Naturally, one obtains the sequence $(t_k)_{k\in\mathbb{N}}$ by ordering the elements of all event series $(t^j_k)_{k\in\mathbb{N}}$ for all agents $j\in\lbrace1,\dots,N\rbrace$ in an increasing fashion.
In addition, let us denote the set of indices $k$ of $(t_k)_{k\in\mathbb{N}}$ which refer to arrived packets by $\mathcal{A}$.
Then, the logic above also applies for the delay series $(d^j_k)_{k\in\mathcal{A}^j}$ and $(d_k)_{k\in\mathcal{A}}$ in the sense that each $d^j_k$ corresponds to a $t^j_{k}$ such that we can utilize the same ordering to obtain $(d_k)_{k\in\mathcal{A}}$.

\section{Cost Analysis for Triggering Schemes}\label{sec:ttc_etc}

In this section, we introduce the triggering schemes TTC and ETC and derive the respective cost according to \eqref{eq:cost}.

\subsection{General Cost Analysis}\label{sec:analysis}
First, we state two useful facts about the considered problem.
Similar to \cite{Rabi2009,Meister2022}, we find:
\begin{fact}\label{fact:0T}
	If the inter-event times $(t_{k+1}-t_k)_{k\in\mathbb{N}}$ are independent and identically distributed, the cost can be computed by considering the time interval between two successful transmissions $\lbrack t^\mathrm{s}_k,t^\mathrm{s}_{k+1}\rbrack$:
	\begin{equation*}
		J = \frac{\E{\frac{1}{2}\sum_{i,j=1}^{N} \int_{t^\mathrm{s}_k}^{t^\mathrm{s}_{k+1}} \left(x_i(t)-x_j(t)\right)^2\diff t}}{\E{t^\mathrm{s}_{k+1}-t^\mathrm{s}_k}},
	\end{equation*}
	where $(t^\mathrm{s}_k)_{k\in\mathbb{N}}=(t_k)_{k\in\mathcal{A}}$ are determined by the triggering schemes introduced in Sections~\ref{sec:TT} and \ref{sec:ET}.
\end{fact}
\begin{pf}
	The proof is omitted due to space limitations.
	It uses similar arguments as in \cite[Fact~1]{Meister2022}.
\end{pf}

\begin{fact}[\cite{Blind2013}, Lemma~35]\label{fact:ETs-ET}
	Let the expected time between two transmission attempts be $\E{T}$ and the packet loss probability be $p$.
	Then, the expected time between two successful transmissions $\E{T^\mathrm{s}}$ is
	\begin{equation*}
		\E{T^\mathrm{s}} = \frac{\E{T}}{1-p}.
	\end{equation*}
\end{fact}

Subsequently, we will use these facts to show that the cost can be separated into components induced by the system behavior without network effects, the transmission delays and packet loss.
Let us first analyze the impact of transmission delays on the performance.
\begin{lem}\label{lem:delay_cost}	%
	Suppose agents \eqref{eq:agent} are controlled by the impulsive input \eqref{eq:input} and the inter-event times $(t_{k+1}-t_k)_{k\in\mathbb{N}}$ are independent and identically distributed.
	Then, the cost \eqref{eq:cost} can be separated into
	\begin{equation}
		J = \bar{J} + N(N-1)\E{d_k}, \quad k\in\mathcal{A}
	\end{equation}
	where $\bar{J}$ refers to the cost without considering transmission delays $(d_k)_{k\in\mathcal{A}}$, i.e., $J$ given $d_k=0$ for all $k\in\mathcal{A}$.
\end{lem}
\begin{pf}
	Since the inter-event times are independent and identically distributed, we can utilize Fact~\ref{fact:0T} to analyze the cost in \eqref{eq:cost}.
	Let us mark all variables neglecting effects of delays with a bar, e.g., $x_i(t)$ refers to the state of agent $i$ including the effect of delays $(d_k)_{k\in\mathcal{A}}$ and $\bar{x}_i(t)$ denotes the same state under the assumption $d_k=0$ for all $k\in\mathcal{A}$.

	Together with Fact~\ref{fact:0T}, the cost can be rewritten as
	\begin{multline*}
		J = \frac{1}{2\E{t^\mathrm{s}_{k+1}-t^\mathrm{s}_k}}\sum_{i,j=1}^{N} \left(\E{\int_{t^\mathrm{s}_k+d^\mathrm{s}_k}^{t^\mathrm{s}_{k+1}} (\bar{x}_i(t)-\bar{x}_j(t))^2 \diff t}\right.\\
		+\left.\E{\int_{t^\mathrm{s}_k}^{t^\mathrm{s}_k+d^\mathrm{s}_k} ((\bar{x}_i(t)+\bar{x}_i(t^\mathrm{s}_k))-(\bar{x}_j(t)+\bar{x}_j(t^\mathrm{s}_k)))^2 \diff t}\right),
	\end{multline*}
	where we utilized the fact that for the transmitting agent $i\in\lbrace 1,\dots,N\rbrace$, the relation $x_i(t)=\bar{x}_i(t) \;\forall t\in\lbrack t^\mathrm{s}_k,t^\mathrm{s}_{k+1}\rbrack$ holds while for all other agents $j\in\lbrace 1,\dots,N\rbrace\backslash\lbrace i\rbrace$, the following equality is fulfilled
	\begin{equation*}
		x_j(t) = \begin{cases}
			\bar{x}_j(t) + (\bar{x}_j(t^\mathrm{s}_k)-\bar{x}_i(t^\mathrm{s}_k))\; & \forall t\in\lbrack t^\mathrm{s}_k,t^\mathrm{s}_k+d^\mathrm{s}_k\rbrack,     \\
			\bar{x}_j(t)\;                                                         & \forall t\in\lbrack t^\mathrm{s}_k+d^\mathrm{s}_k,t^\mathrm{s}_{k+1}\rbrack.
		\end{cases}
	\end{equation*}
	Rearranging terms yields
	\begin{multline*}
		J = \frac{1}{2 \E{t^\mathrm{s}_{k+1}-t^\mathrm{s}_k}}\sum_{i,j=1}^{N} \left(\E{\int_{t^\mathrm{s}_k}^{t^\mathrm{s}_{k+1}} (\bar{x}_i(t)-\bar{x}_j(t))^2 \diff t}\right.\\
		+2\E{\int_{t^\mathrm{s}_k}^{t^\mathrm{s}_k+d^\mathrm{s}_k} (\bar{x}_i(t)-\bar{x}_j(t))(\bar{x}_i(t^\mathrm{s}_k)-\bar{x}_j(t^\mathrm{s}_k)) \diff t}\\
		+\left.\E{\int_{t^\mathrm{s}_k}^{t^\mathrm{s}_k+d^\mathrm{s}_k} (\bar{x}_i(t^\mathrm{s}_k)-\bar{x}_j(t^\mathrm{s}_k))^2 \diff t}\right),
	\end{multline*}
	where the middle term vanishes since $\bar{x}_i(t)-\bar{x}_j(t)$ and $\bar{x}_i(t^\mathrm{s}_k)-\bar{x}_j(t^\mathrm{s}_k)$ are independent and $\E{\bar{x}_i(t)-\bar{x}_j(t)}=0$ for all $t\in\lbrack t^\mathrm{s}_k,t^\mathrm{s}_k+d^\mathrm{s}_k\rbrack$.
	Abbreviating the first term as $\bar{J}$ and computing the last one leads to
	\begin{align*}
		J & = \bar{J} + \frac{1}{2}\sum_{i,j=1}^{N} \frac{\E{(\bar{x}_i(t^\mathrm{s}_k)-\bar{x}_j(t^\mathrm{s}_k))^2} \cdot \E{d^\mathrm{s}_k}}{\E{t^\mathrm{s}_{k+1}-t^\mathrm{s}_k}} \\
		  & = \bar{J} + \frac{1}{2}\sum_{\substack{i,j=1:\\ i\neq j}}^{N} \frac{2\E{T^\mathrm{s}} \cdot \E{d^\mathrm{s}_k}}{\E{T^\mathrm{s}}}                                          \\
		  & = \bar{J} + N(N-1) \E{d_k},	\quad k\in\mathcal{A}
	\end{align*}
	which results from $d^\mathrm{s}_k$ and $\bar{x}_i(t^\mathrm{s}_k)$ being independent for all $i\in\lbrace1,\dots,N\rbrace$.
	\hfill\hfill\qed
\end{pf}

We have therefore proven that the cost can be separated into a component induced by the transmission delays and one that neglects them.
Let us refine this separation further by considering the cost induced by packet loss.
\begin{lem}\label{lem:loss_cost}
	Suppose agents \eqref{eq:agent} are controlled by the impulsive input \eqref{eq:input} and the inter-event times $(t_{k+1}-t_k)_{k\in\mathbb{N}}$ are independent and identically distributed.
	Moreover, let $p$ be the packet loss probability.
	Then, the cost \eqref{eq:cost} can be separated into
	\begin{equation*}
		J = \dbar{J} + N(N-1) \left(\frac{p\,\E{T}}{1-p}+\E{d_k}\right),	\quad k\in\mathcal{A}
	\end{equation*}
	with the cost $\dbar{J}$ neglecting packet loss and transmission delays as well as the expected inter-event time $\E T$, namely the time between two transmission attempts.
\end{lem}
\begin{pf}
	Applying Fact~\ref{fact:0T} and Lemma~\ref{lem:delay_cost}, we analyze the interval between two successful packet transmissions in the scenario with packet loss but without delays.
	Since we are neglecting delays, we may analyze the first interval between two successful packet transmissions $[0,T^\mathrm{s}]=[0,t^\mathrm{s}_1]$ to simplify notation
	\begin{equation*}
		\bar{J} = \frac{\E{\int_{0}^{T^\mathrm{s}} \frac{1}{2}\sum_{i,j=1}^{N}\left(\bar{x}_i(t)-\bar{x}_j(t)\right)^2\diff t}}{\E{T^\mathrm{s}}}.
	\end{equation*}
	We can express the numerator as
	\begin{align*}
		    & \E{\int_{0}^{T^\mathrm{s}} \frac{1}{2}\sum_{i,j=1}^{N}\left(\bar{x}_i(t)-\bar{x}_j(t)\right)^2\diff t}                                                      \\
		={} & \sum_{m=1}^{\infty}(1-p)p^{m-1} \E{\int_{0}^{t_m} \frac{1}{2}\sum_{i,j=1}^{N}\left(\bar{x}_i(t)-\bar{x}_j(t)\right)^2\diff t}                               \\
		={} & \frac{1}{2}\sum_{i,j=1}^{N} \frac{1-p}{p} \sum_{m=1}^{\infty}p^{m} \sum_{k=1}^{m} \E{\int_{t_{k-1}}^{t_k} \left(\bar{x}_i(t)-\bar{x}_j(t)\right)^2\diff t},
	\end{align*}
	where $t_m$ denotes the time instant of the $m$-th transmission attempt.
	Computing the expected value from the previous equation for $i\neq j$ yields
	\begin{align*}
		    & \E{\int_{t_{k-1}}^{t_k} \left(\bar{x}_i(t)-\bar{x}_j(t)\right)^2\diff t}                                                                    \\
		={} & \left(\E{\bar{x}_i(t_{k-1})^2}+\E{\bar{x}_j(t_{k-1})^2}\right)\E{t_k-t_{k-1}}                                                               \\
		+{} & \E{\int_{t_{k-1}}^{t_k}\left(\left(\bar{x}_i(t)-\bar{x}_i(t_{k-1})\right) - \left(\bar{x}_j(t)-\bar{x}_j(t_{k-1})\right)\right)^2 \diff t},
	\end{align*}
	where we utilized the fact that $\bar{x}_i(t)$ and $\bar{x}_j(t)$ are independent for $i\neq j$ and all $t\geq0$.
	Moreover, the last term considers the state evolution without packet loss and transmission delays since the agent states are reset to consensus at every $t_{k-1}$.
	Thus, we have
	\begin{align*}
		    & \E{\int_{t_{k-1}}^{t_k} \left(\bar{x}_i(t)-\bar{x}_j(t)\right)^2\diff t}                       \\
		={} & 2\E{t_{k-1}}\E{T} + \E{\int_{t_{k-1}}^{t_k} \left(\dbar{x}_i(t)-\dbar{x}_j(t)\right)^2\diff t} \\
		={} & 2(k-1)\E{T}^2 + \E{\int_{t_{k-1}}^{t_k} \left(\dbar{x}_i(t)-\dbar{x}_j(t)\right)^2\diff t},
	\end{align*}
	where we abbreviate the state evolution without transmission delays and packet loss as $\dbar{x}_i(t)$ for all $i\in\lbrace1,\dots,N\rbrace$.
	Combining all derivations with Fact~\ref{fact:ETs-ET} yields the cost
	\begin{align*}
		\bar{J} & =\frac{1}{2\E{T}} \sum_{\substack{i,j=1:\\i\neq j}}^{N} \frac{(1-p)^2}{p} \sum_{m=1}^{\infty}p^{m}                         \\
		        & \quad\cdot \left( 2\E{T}^2\frac{m(m-1)}{2} + m\E{\int_{0}^{t_1} \left(\dbar{x}_i(t)-\dbar{x}_j(t)\right)^2\diff t} \right) \\
		        & = \frac{1}{2} \sum_{\substack{i,j=1:\\i\neq j}}^{N} \left(\E{T}\left(\frac{1+p}{1-p}-1\right)\right.                       \\
		        & \qquad\qquad\quad \left.+ \frac{1}{\E T} \E{\int_{0}^{T} \left(\dbar{x}_i(t)-\dbar{x}_j(t)\right)^2\diff t}\right)         \\
		        & = N(N-1)\frac{p\,\E{T}}{1-p} + \dbar{J},
	\end{align*}
	where we utilized closed-form expressions for low-order polylogarithms.
	Applying Lemma~\ref{lem:delay_cost} finishes the proof.
	\hfill\qed
\end{pf}

\subsection{Time-Triggered Control}\label{sec:TT}
In TTC, transmission events are scheduled periodically, i.e., with a constant inter-event time $\TTT = t_{k+1}-t_{k} = \mathrm{const.}$ for all $k\in\mathbb{N}$.
At each event, one agent broadcasts its state to all other agents.
If the packet with state information is not lost, it arrives after a delay and allows all agents to apply control input \eqref{eq:input}.
The transmitting agent can be chosen according to an arbitrary scheme as the cost is not influenced by that choice in our setup.

Deploying this triggering scheme in the considered setup allows us to arrive at the following theorem.
\begin{thm}\label{thm:cost_TT}
	Let agents \eqref{eq:agent} be controlled by the impulsive input \eqref{eq:input} with constant inter-event times $\TTT$.
	Then, the cost \eqref{eq:cost} is given by
	\begin{equation*}
		J_{\mathrm{TT}} = N(N-1) \left( \frac{\TTT}{2} + \frac{p\,\TTT}{1-p} + \E{d_k} \right).
	\end{equation*}
\end{thm}
\begin{pf}
	The result follows from combining Lemma~\ref{lem:loss_cost} with \cite[Theorem~1]{Meister2022} showing
	\begin{equation*}
		\dbar{J}_{\mathrm{TT}} = N(N-1) \frac{\TTT}{2}.
	\end{equation*}
\end{pf}

\begin{rem}
	Note that the structure of the cost is similar to the one in \cite[Theorem~4]{Blind2013} but scaled by the number of agent pairs.
\end{rem}

\subsection{Event-Triggered Control}\label{sec:ET}
In ETC, a transmission is initiated by an agent if a triggering condition is fulfilled.
The triggering condition should be chosen such that it indicates communication necessity in the setting at hand.
As in \cite{Meister2022}, we define
\begin{equation}\label{eq:ET_cond}
	\lvert x_i(t)-x_i(t_{\hat{k}}) \rvert \geq \Delta,
\end{equation}
as the triggering condition where $\hat{k} = \max \left\lbrace k\in\mathbb{N} \mid t_{k} \leq t \right\rbrace$ and $\Delta>0$.
In a distributed setup, it is quite common to compare the deviation of the current state $x_i(t)$ of agent $i$ from its state at the last event $x_i(t_{\hat{k}})$ to a threshold $\Delta$.
This is due to the fact that each agent can check this condition locally, see, e.g., \cite{Dimarogonas2009}.
Since each agent's state contributes equally to the cost, we utilize the same threshold $\Delta$ for all agents.
Note that we use a triggering condition analogous to \cite{Astrom2002}, \cite{Rabi2009}, \cite{Blind2013} but ours is of a distributed nature.
To deploy \eqref{eq:ET_cond} in a distributed setup, we additionally need the following assumption.

\begin{assum}\label{assum:perf_channel_ETC}
	Each agent is able to indicate instantaneously to all other agents when a packet transmission is started.
	This information is broadcasted to the other agents without loss or delay.
\end{assum}

In practice, this can for example be achieved by reserving a certain frequency range of the network bandwidth only for indicating transmission time instants with a broadcast.
Note that the transmission of one bit is sufficient to indicate the transmission time instant to all other agents.
Thus, the reserved bandwidth can be rather small while still enabling close to zero transmission times and, thus, close to zero packet loss probabilities.

In ETC, we can express the inter-event time as a stopping time $\TET = \inf\lbrace t>0 \mid \exists i\in\lbrace1,\dots,N\rbrace: \lvert x_i(t)\rvert = \Delta \rbrace$, leveraging the argument from Fact~\ref{fact:0T}.
Therefore, the inter-event time is a stochastic variable.
Note that the triggering condition \eqref{eq:ET_cond} with Assumption~\ref{assum:perf_channel_ETC} renders the inter-event times of the ETC scheme independent and identically distributed.
This allows us to apply Lemma~\ref{lem:loss_cost}, but the derivation of a closed form expression for the cost $\dbar{J}_\mathrm{ET}$ is not possible.
Nonetheless, we can utilize the result from \cite[Theorem~2]{Meister2022} to determine the relationship to the TTC cost.

\begin{lem}[\cite{Meister2022}]\label{lem:cost_ET}
	Let agents \eqref{eq:agent} be controlled by the impulsive input \eqref{eq:input} with inter-event times $\TET = \inf\lbrace t>0 \mid \exists i\in\lbrace1,\dots,N\rbrace: \lvert x_i(t)\rvert = \Delta \rbrace$.
	Then, there exists an $N_0$ such that for all $N\geq N_0$,
	\begin{equation*}
		\dbar{J}_\mathrm{ET} > \dbar{J}_\mathrm{TT}(\TTT=\E{\TET}),
	\end{equation*}
	where we denote by $\dbar{J}_\mathrm{TT}(\TTT=\E{\TET})$ the cost under constant inter-event times $\TTT=\E{\TET}$ without considering packet loss or transmission delays.
\end{lem}

This result points out that ETC is not necessarily superior to TTC in this MAS setup under the assumption that network effects can be neglected.
In order to include packet loss and transmission delay effects in the comparison, we need to determine packet loss probability and expected delay based on the deployed MAC protocols.

\section{Medium Access Control Protocols}\label{sec:mac}

The packet loss probability $p$ and the expected transmission delay $\E{d_k}$ depend on the MAC protocol.
We will analyze TTC and ETC under simple MAC protocols forming the basis of complex ones.
As \cite{Blind2013}, we examine Time Division Multiple Access (TDMA) for TTC and pure ALOHA (\cite{Abramson1970}) for ETC.

The TDMA protocol belongs to the group of deterministic MAC protocols which reserve the shared resource a priori in order to prevent packet loss.
It is therefore only applicable to TTC.
The pure ALOHA protocol belongs to the class of contention-based MAC protocols.
They do not schedule the access to the shared network for each agent a priori at the risk of packet collisions.

Applying a contention-based MAC protocol to the TTC scheme will not result in a performance improvement when compared to TDMA.
As argued by \cite{Blind2013}, Section~5.2, a properly designed contention-based protocol applied to TTC can achieve TDMA performance at best because of the deterministic nature of the transmission attempts.
We will therefore analyze TDMA for TTC while evaluating the contention-based protocol for ETC.

\subsection{Time Division Multiple Access}\label{sec:TDMA}
In the TDMA protocol, the transmission events are assigned to the agents in advance.
Therefore, packet collisions are completely prevented as long as the inter-event time $\TTT$ is larger than the transmission time $\tau$, i.e.,
\begin{equation*}
	p_\mathrm{TDMA} = \begin{cases}
		0, \quad & \TTT \geq \tau,     \\
		1, \quad & \mathrm{otherwise},
	\end{cases}
\end{equation*}
and the expected delay equals the transmission time, i.e., $\E{d_{\mathrm{TDMA},k}}=\tau$.

\begin{thm}\label{thm:ttc_tdma}
	Applying TTC with the TDMA protocol to agents \eqref{eq:agent} with input \eqref{eq:input} yields the normalized cost
	\begin{equation*}
		\tilde{J}_{\mathrm{TT,TDMA}} = \frac{1}{2\rho} + 1\quad\forall \rho\in(0,1\rbrack,
	\end{equation*}
	where we normalized by $\tau N(N-1)$ and $\rho = \tau/\TTT$ denotes the network load.
\end{thm}
\begin{pf}
	The result follows from utilizing $p_\mathrm{TDMA}$ and $\E{d_{\mathrm{TDMA},k}}$ in the cost expression in Theorem~\ref{thm:cost_TT} as well as normalizing by $\tau N(N-1)$.
\end{pf}

For a finite cost, the maximum network load is $\rho=1$ since all packets are lost beyond that load.

\subsection{Pure ALOHA}
In the pure ALOHA protocol, each agent has access to the medium at all times.
If a packet is sent while another one is in transmission, both packets are lost.
While $\E{d_{\mathrm{PA},k}} = \tau$ holds for the transmission delay, we can establish the following lemma to compute the packet loss probability.
\begin{lem}\label{lem:p_PA}
	Suppose agents \eqref{eq:agent} are controlled by the input \eqref{eq:input} with ETC triggering condition \eqref{eq:ET_cond}.
	Then, the packet loss probability under the pure ALOHA protocol is
	\begin{equation*}
		p_{\mathrm{PA}} = 1-\Prob(\TET>\tau)^2,
	\end{equation*}
	where, with $\rho = \tau/\E\TET$, we have
	\begin{multline*}
		\Prob(\TET>\tau) = \left(\frac{4}{\pi} \sum_{n=0}^{\infty} \frac{(-1)^n}{(2n+1)} \right. \\
		\left. \cdot\exp\!\left(-\frac{(2n+1)^2 \pi^2\E{\TET \mid \Delta=1}}{8} \, \rho\right) \right)^N,
	\end{multline*}
\end{lem}
\begin{pf}
	Analogously to \cite{Sant1980}, we can compute the loss probability as
	\begin{equation*}
		p = 1 - (1 - \Prob(\TET\leq\tau))^2 = 1-\Prob(\TET>\tau)^2,
	\end{equation*}
	where a packet is lost if it is sent while the previous one is still transmitted or the following packet is sent while the considered one is still transmitted itself.

	Let us define $T_{0,a}^i := \min\lbrace T_0^i,T_a^i\rbrace$, where $T_y^i := \inf\lbrace t>0\mid v_i(t)=y\rbrace$ is the first hitting time of level $y\in\mathbb{R}$ for a Brownian motion $v_i(t)$ starting at $v_{0,i}\in\mathbb{R}$.
	In \cite[(7.15)]{Moerters2010}, it is shown that for $0<v_{0,i}<a_i$,
	\begin{align*}
		\Prob_{v_{0,i}}( T_{0,a_i}^i > t ) ={} & \Prob_{v_{0,i}}( v_i(s)\in(0,a_i) \;\forall s\in\lbrack0,t\rbrack)                                     \\
		={}                                    & \frac{4}{\pi} \sum_{n=0}^\infty \frac{1}{2n+1} \exp\left( - \frac{(2n+1)^2 \pi^2}{2a_i^2} \, t \right) \\
		                                       & \cdot\sin\left( \frac{(2n+1)\pi v_{0,i}}{a_i}\right).
	\end{align*}
	Since we are analyzing multiple independent Brownian motions and their minimum first hitting time for one of the fixed interval limits $0$ and $a_i$, we are interested in
	\begin{equation*}
		\Prob_{v_{0,1},\dots,v_{0,N}}(T_{0,a_1}^1 \wedge\ldots\wedge T_{0,a_N}^N > t) = \prod_{i=1}^{N}\Prob_{v_{0,i}}(T_{0,a_i}^i > t),
	\end{equation*}
	where we utilized that the Brownian motions are mutually independent and $\wedge$ computes the minimum of two quantities.
	Utilizing that all Brownian motions are identically distributed, we have $\Prob(\TET>\tau) = \Prob_{v_{0,1}}(T_{0,a_1}^1 > \tau)^N$ for $a_i = 2\Delta$ and $v_{0,i} = a_i/2$ for all $i\in\lbrace1,\dots,N\rbrace$.
	This yields
	\begin{align*}
		\Prob(\TET>\tau) & = \Prob_{\Delta}(T_{0,2\Delta}^1 > \tau)^N                                                                                             \\
		                 & = \left( \frac{4}{\pi} \sum_{n=0}^\infty \frac{(-1)^n}{2n+1} \exp\left( - \frac{(2n+1)^2 \pi^2}{8 \Delta^2} \, \tau \right) \right)^N.
	\end{align*}
	Plugging in $\E\TET = \Delta^2 \E{\TET\mid\Delta=1}$ from \cite[Fact~3]{Meister2022} and defining the network load $\rho = \tau/\E\TET$ gives us the desired result.
	\hfill\hfill\qed
\end{pf}

After quantifying the expected transmission delay and packet loss probability for ETC with the pure ALOHA protocol, we arrive at the respective normalized cost.

\begin{thm}\label{thm:etc_pa}
	Applying ETC with triggering rule \eqref{eq:ET_cond} and the pure ALOHA MAC protocol to agents \eqref{eq:agent} with input \eqref{eq:input} yields the cost, normalized by $\tau N(N-1)$,
	\begin{multline*}
		\tilde{J}_{\mathrm{ET,PA}} = \frac{\dbar{J}_\mathrm{ET}(\Delta=1)}{N(N-1)\E{\TET\mid\Delta=1}}\cdot\frac{1}{\rho} \\
		+ \frac{p_{\mathrm{PA}}}{1-p_{\mathrm{PA}}} \cdot\frac{1}{\rho} + 1\quad\forall \rho\in(0,\infty),
	\end{multline*}
	with network load $\rho = \tau/\E\TET$.
\end{thm}
\begin{pf}
	The result follows from combining Lemma~\ref{lem:loss_cost} with \cite[Fact~3]{Meister2022} and $\E{d_{\mathrm{PA},k}} = \tau$ as well as normalizing by $\tau N(N-1)$.
\end{pf}

Although we do not arrive at a fully explicit expression for the cost since $\dbar{J}_\mathrm{ET}(\Delta=1)$ and $\E{\TET\mid\Delta=1}$ are not known explicitly from our derivation, we can still deduce some important properties from Theorem~\ref{thm:etc_pa}.
For example, combining the findings from Lemma~\ref{lem:cost_ET} and \cref{thm:ttc_tdma,thm:etc_pa} allows us to deduce that there exists a number of agents $N_0$ beyond which TTC with TDMA certainly outperforms ETC with pure ALOHA regardless of the network load.
In contrast to \cite{Meister2022}, we could confirm that this also holds when incorporating network effects under the considered MAC protocols and for any network load.
Moreover, as $p_\mathrm{PA}$ is monotonically increasing from 0 to 1 on the range $\rho\in(0,\infty)$, TTC with TDMA outperforms ETC with pure ALOHA for sufficiently high network loads $\rho\in(0,1\rbrack$ regardless of the number of agents.
This is due to the following simple argument:
$\Prob(\TET>\tau \mid \rho=1) = \Prob(\TET>\E{\TET}) = 1/2$ by definition of the expected value.
Thus, $p_\mathrm{PA}(\rho=1) = 1-\Prob(\TET>\E{\TET})^2 = 3/4$ while $\tilde{J}_{\mathrm{ET,PA}} > \tilde{J}_{\mathrm{TT,TDMA}}$ for $p_\mathrm{PA} > 1/3$.
This finding is particularly relevant because it demonstrates network-induced performance limitations for ETC compared to TTC.
In addition, we will utilize the simulation results and method from \cite{Meister2022} to compute $\tilde{J}_{\mathrm{ET,PA}}$ for various network loads and numbers of agents, and contrast it to $\tilde{J}_{\mathrm{TT,TDMA}}$ in the next section.

\section{Numerical Evaluation}\label{sec:sim}

In this section, we complement our theoretical findings with a numerical evaluation of the normalized costs $\tilde{J}_{\mathrm{ET,PA}}$ and $\tilde{J}_{\mathrm{TT,TDMA}}$ for varying network loads $\rho$.
For the normalized ETC cost, we rely on the simulation results from \cite{Meister2022} to quantify $\E{\TET\mid\Delta=1}$ and $\dbar{J}_\mathrm{ET}(\Delta=1)$ for various $N$.
For a fair comparison of TTC and ETC, we require $\TTT=\E{\TET}$, i.e., both triggering schemes attempt transmissions with the same average rate.
Note that due to normalization, we arrive at only one normalized cost curve for $\tilde{J}_{\mathrm{TT,TDMA}}$ while $\tilde{J}_{\mathrm{ET,PA}}$ varies with the number of agents $N$ in the network.

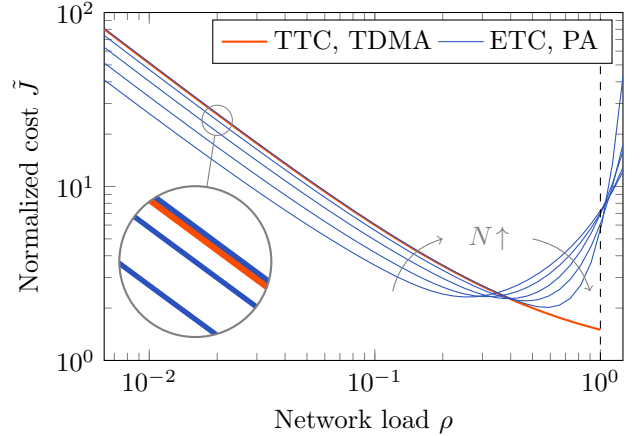
\begin{figure}
	\centering
	\input{img/plot_cost.tex}
	\vspace*{-8pt}
	\caption{Normalized cost of TTC with TDMA and ETC with pure ALOHA for $N\in\lbrace2,3,6,12,72\rbrace.$}
	\label{fig:cost}
\end{figure}

The resulting normalized cost over the network load is shown in Fig.~\ref{fig:cost}.
Although not shown by our theoretical results, we observe that, for low numbers of agents $N$ and low network loads $\rho$, the ETC scheme has a clear performance advantage over TTC.
At low network loads or sufficiently high performance levels, this confirms the intuition of ETC providing the same performance level as TTC while saving communication resources.
The performance advantage of ETC shrinks with an increasing number of agents in the network.
Furthermore, the ETC scheme is outperformed by TTC for larger numbers of agents $N$ regardless of the network load and for large network loads $\rho\in(0,1\rbrack$ regardless of the number of agents.
The latter is due to the fact of uncoordinated medium access and, thus, a higher packet loss probability in the case of ETC and pure ALOHA.
We have therefore demonstrated that network effects can lead to additional performance degradation for ETC compared to TTC.
Thus, we have shown that \cite[Theorem~2]{Meister2022} holds independently of the network load when incorporating network effects with the considered MAC protocols.
Moreover, the numerical evaluations reveal that for higher network loads, performance advantages of ETC for small agent numbers are outweighed by the negative impact of network effects in our setting.

\section{Conclusion}\label{sec:conclusion}

In this work, we analyzed the performance of TTC and ETC in a MAS consensus setup with single-integrator agents including transmission delays and packet loss due to imperfect communication.
We thereby extended the performance comparison in \cite{Meister2022} by the consideration of network effects.
Our work highlights that TTC might outperform ETC in this particular setup for a sufficiently large number of agents or high network loads.
We have shown that the former result, also stated in \cite{Meister2022}, still holds when network effects are taken into account for the considered setup and MAC protocols.
In addition, the found performance advantage of TTC for high network loads demonstrates that network effects are non-negligible for performance evaluations of ETC and TTC.
In conclusion, this work demonstrates for the analyzed setup that a performance-oriented design decision on ETC versus TTC under consideration of network effects can be more complex than for non-cooperative NCS.
This is due to the fact that the performance relationship turns out to depend on the network load as well as the number of agents for the considered setting.
In future work, we plan to analyze more MAC protocols such as slotted ALOHA and their impact on the induced performance of TTC and ETC.
Moreover, the generalization to a broader class of communication topologies will unfold many additional research topics such as the applicability and implications of various MAC protocols in other network scenarios.

\begin{ack}
	We thank Frank Aurzada from the Technical University of Darmstadt for the fruitful discussions.
\end{ack}

\renewcommand*{\bibfont}{\scriptsize}
\bibliography{references}   %

\appendix
\end{document}

%% file: img/plot_cost.tex
\usetikzlibrary{spy}
\includegraphics{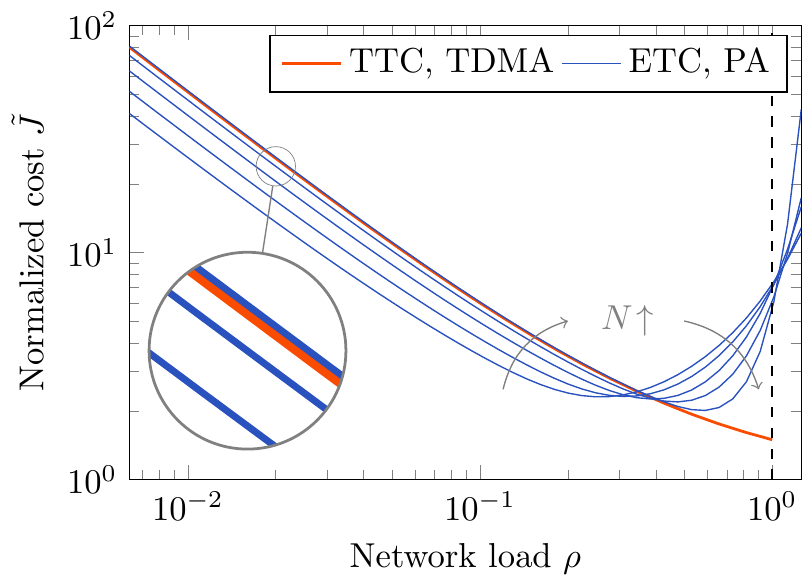}